\documentclass{article}
\usepackage[utf8]{inputenc}

\def\includegraphics{}
\usepackage{graphicx}
\usepackage{subcaption}
\usepackage[normalem]{ulem}
\usepackage{authblk}
\usepackage{amsmath,amssymb}
\usepackage{lipsum}
\usepackage{hyperref}
\usepackage[numbers,sort]{natbib}
\usepackage[labelformat=empty]{subcaption}
\usepackage[figurename=Fig.]{caption}

\title{The power of dynamic social networks to predict individuals' mental health}
\author[1]{Shikang Liu}
\author[2]{David Hachen}
\author[3]{Omar Lizardo}
\author[1]{Christian Poellabauer}
\author[1]{Aaron Striegel}
\author[1,4,5,*]{Tijana Milenkovi\'{c}}

\affil[1]{Department of Computer Science and Engineering, University of Notre Dame}
\affil[2]{Department of Sociology, University of Notre Dame}
\affil[3]{Department of Sociology, University of California, Los Angeles}
\affil[4]{Eck Institute for Global Health, University of Notre Dame}
\affil[5]{Interdisciplinary Center for Network Science and Applications (iCeNSA), University of Notre Dame}
\affil[*]{Corresponding author (email: tmilenko@nd.edu)}

\date{ }

\begin{document}

\maketitle

\begin{abstract}
Precision  medicine has received attention both in and  outside  the  clinic. We focus on the latter, by exploiting the relationship between individuals' social interactions and their mental health to develop a predictive model of one's likelihood to be depressed or anxious from rich dynamic social network data. To our knowledge, we are the first to do this. Existing studies differ from our work in at least one aspect: they do not model social interaction data as a network; they do so but analyze static network data; they examine  ``correlation'' between social networks and health but without developing a predictive model; or they study other individual traits but not mental health. In a systematic and comprehensive evaluation, we show that our predictive model that uses dynamic social network data is superior to its static network as well as non-network equivalents when run on the same data. Supplementary material for this work is available at \url{https://nd.edu/~cone/NetHealth/PSB_SM.pdf}.

\end{abstract}

%%%%%%%%%%%%%%%%%%%%%%%%%%%%%%%%%%%%%%%%%%%%%%%%%%%%%%%%%%%%%%%%%%%%%%%%%%%%%%%%%%%%%%%%%%%%%%%%%%%%%%%%%%%%%%%%%%%%%%%%%%%%%%%%%%%%%%%%%%%%%%%%%%%%%%%%%%%%%%%%%%%%%%%%%%%%%%%%%%%%%%%%%%%%%%%%%%%%%%%%%%%%%%%%%%%%%%%%%%%%%%%%%%%%%%%%%%%%%%%%%%%%%%%%%%%%%%%%%%%%%%%%

\section{Introduction} \label{intro}

Precision medicine, i.e., giving personalized health-related recommendations to individuals, has received attention both in the clinic via -omics data collection technologies \cite{malod2018precision,gligorijevic2016integrative} and outside the clinic via mobile health tracking devices such as smartphones or wearable sensors, online social media behavioral data, or other non-traditional personal health data resources \cite{liu2019heterogeneous,latkin2015social,perkins2015social,pastor2015epidemic,valente2017appraisal,cobb2010social}. In this study, we focus on the role of individuals' social interactions on their health. For example, individuals' health-related traits, such as obesity, smoking, depression, physical activities, heart rates, self-perceived health, happiness, or stress can spread through social interactions \cite{christakis2007spread,christakis2008collective,rosenquist2011social} or are correlated with the individuals' positions (e.g., centralities) in their social network \cite{youm2014social,liu2018network,lin2019social,schafer2011health,haas2010health,meng2016interplay}.

In this study, we focus on \textit{mental} health, specifically depression and anxiety, because they are critical public health issues affecting millions of individuals worldwide \cite{berk2011does,aldarwish2017predicting,mcgorry2015early}. In particular, we focus on developing a machine learning model for predicting individuals' mental health conditions, whose input is a feature vector for each individual extracted from a given data source, and whose output is a predicted likelihood of an individual being depressed or anxious.
%, which is computed in a supervised classification manner based on the ``similarity'' of the given individual's feature to the features of other individuals and the knowledge of which of the other individuals are depressed or anxious. 

Existing studies that propose machine learning models for predicting mental health can be divided into non-network ones and network ones. Non-network studies rely on smartphone usage data such as incoming and outgoing call frequency, wearable sensor data such as physical activity (e.g., step count), or online social media behavioral data such as text content on social media platforms \cite{mohr2017personal, glenn2014new,guntuku2017detecting,bogomolov2014daily,wongkoblap2017researching,de2013predicting}. These studies typically extract features from such data in its raw form and input those features into off-the-shelf classifiers to make mental health predictions. On the other hand, network studies either use the same data as non-network studies but first process the raw data into a network that captures relationships between entities (e.g., by linking two individuals if their physical activity profiles are similar), or they use explicit social network data (e.g., friendships between individuals). Then, they extract network-based features (e.g., an individual's centralities in the network) and input those features into off-the-shelf classifiers \cite{wang2013improved,lin2019social,meng2016interplay,liu2018network}. Since networks capture complex wirings between entities, network studies are expected to be advantageous. Indeed, in our previous study \cite{liu2019heterogeneous}, we showed that network analysis of rich social data  originating from the NetHealth study \cite{purta2016experiences, faust2017exploring} was more accurate than non-network analysis of \textit{the same data} in the task of predicting individuals' mental health.

In more detail, this previous study of ours \cite{liu2019heterogeneous} is the most comprehensive work to date on predicting individuals' mental health in terms of both data size (the number of considered individuals) and data heterogeneity (the number of considered data types). Namely, we leveraged the rich NetHealth data set containing individuals' social interaction data (SMS communications), health-related behavioral data (physical activity and sleep duration) and a variety of individuals' trait data (personality traits, social status, physical health, and well-being) to predict mental health conditions \cite{liu2019heterogeneous}. We integrated such data into a heterogeneous information network (HIN). Then, we modeled the problem of mental health prediction from our HIN in a novel manner, as applying to the HIN a popular paradigm of a recommender system (RS), which is typically used to predict the preference that an individual would give to an item (e.g., a movie or book). In our case, the items were the individuals' different mental health states. RS gives us an intuitive yet powerful way to predict an individual's mental health state (likelihood of being depressed or anxious) by relying on information about both the individual's and her/his neighbors' behaviors and traits. We found that RS produced more accurate predictions than other types of network methods as well as a fairly comparable \textit{non-network} method using the same data. This confirms the power of our HIN-based predictive framework.

The existing RS methods (and other types of HIN methods) work on static network data, meaning that they make predictions from a static network in which nodes and edges do not change with time. Extending them to be able to deal with dynamic network data is non-trivial because it requires novel methodologies to model individuals' dynamic traits and dynamic social interactions \cite{koren2009collaborative}. For these reasons, we constructed our HIN by aggregating dynamic social network data from the considered study time period into a static network.
%, meaning that social interactions at different time points are projected as edges of a single network. 
And while even this led to high mental health prediction accuracy,  temporal information was lost. Including  temporal information could further improve prediction accuracy compared to using static network data. So, in this study, we fairly evaluate the power of using dynamic versus static social network data in the task of predicting mental health. If the former has more predictive power, our HIN-based predictive framework from our previous study \cite{liu2019heterogeneous} could be extended to incorporate dynamic instead of static social network data. Since doing this is non-trivial, as discussed above, it is out of the scope of the current paper and is instead left as future work.

Compared to existing studies on associating social networks and individuals' traits (Table \ref{Table:1}), our study differentiates from each of those in at least one of the following four key aspects: 1) whether the study is network-based or not; 2) if yes, whether the study considers a dynamic network or a static network; 3) whether the study builds machine learning models to predict individuals' traits or ``only'' examines the existence of a potential correlation-like relationship between social networks and traits without making any predictions; 4) whether the study deals with studying mental health or instead it studies other traits.

In more detail, a group of studies built machine learning models to predict mental health, but they are non-network studies \cite{bogomolov2014daily,de2013predicting,glenn2014new,guntuku2017detecting, mohr2017personal,sano2015recognizing,wongkoblap2017researching}, unlike our network study. And as discussed above, we already showed that network studies are superior to non-network  studies in this task \cite{liu2019heterogeneous}. Among network studies (including those that did not necessarily focus on mental health), a majority only explored the existence of a potential  ``correlation''  between individuals' social networks and their traits (e.g., whether individuals who have different personality traits tend to have different centrality values). That is, they did not build any predictive models \cite{bollen2011happiness,christakis2007spread,christakis2008collective,haas2010health,hatzenbuehler2012social,krause2010personality,meng2016interplay,schafer2011health,youm2014social,liu2018network,rosenquist2011social,schaefer2011misery}, which is what we aim to do here. Plus, a majority of these did not focus on mental health, which we do in this study. Of the four network studies that did build predictive models, three used static networks instead of dynamic networks \cite{liu2019heterogeneous,staiano2012friends,wang2013improved}, and thus, they could not compare the predictive power of dynamic versus static networks, which we do in this study. Plus, of these three, one study did not even deal with mental health; instead, it dealt with personality traits \cite{staiano2012friends}. The only study that did build predictive models using dynamic social network data did not ask the research question of our study\textemdash whether using dynamic network data is more predictive than using static network data \cite{lin2019social}. Moreover, while it\cite{lin2019social} studied some health-related traits (e.g., happiness, positive attitude) in addition to some non-health-related traits (e.g., gender, race), it did not study mental health (i.e., depression and anxiety) as we do in our study. Also, to predict individuals' traits, this existing study \cite{lin2019social} used as features centralities of nodes (i.e., individuals) in the dynamic network data. This is a typical strategy in many tasks and domains \cite{meng2016interplay, faisal2014dynamic} and is thus not unique to this existing study \cite{lin2019social}. In our study, on top of also using node centralities, we consider additional and possibly more powerful node features that are based on graphlets \cite{aparicio2018graphlet,milenkovic2008uncovering,hulovatyy2015exploring}. Graphlets are subgraphs, i.e., Lego-like basic building blocks of complex networks. They are state-of-the-art features, especially in dynamic networks \cite{aparicio2018graphlet,milenkovic2008uncovering,hulovatyy2015exploring}. While these features were proposed by others \cite{aparicio2018graphlet,milenkovic2008uncovering,hulovatyy2015exploring}, we are the first of all social-networks-to-health studies to use them, which further distinguishes our work from the related studies. Note that this existing study \cite{lin2019social} was just published, and we became aware of it after we completed all of our analyses and as we were finalizing our paper. So, that study \cite{lin2019social} and our work proposed here can be considered as concurrent pieces of work. 

Hence, in terms of novelty, we are the first ones to develop predictive models of mental health (depression and anxiety) from dynamic social network data and to study whether using dynamic social network data is more predictive than using static social network data.

\begin{table}[h!]
\centering
\resizebox{\textwidth}{!}{\begin{tabular}{ |c|c|c|c|c| } 
 \hline
 & Network-based? & Dynamic network? & Predictive model? & Mental health? \\ 
  \hline
 Our study & \checkmark  & \checkmark & \checkmark &  \checkmark \\ 
 \hline
   \cite{bogomolov2014daily,de2013predicting,glenn2014new,guntuku2017detecting, mohr2017personal,sano2015recognizing,wongkoblap2017researching} & &  & \checkmark & \checkmark\\
   \hline
     \cite{bollen2011happiness,haas2010health,krause2010personality,schafer2011health,youm2014social}  & \checkmark & &  & \\
   \hline
     \cite{hatzenbuehler2012social} & \checkmark & &  & \checkmark \\
 \hline
 \cite{staiano2012friends} & \checkmark & & \checkmark &  \\
   \hline
 \cite{christakis2007spread,christakis2008collective,meng2016interplay} & \checkmark & \checkmark &  &  \\ 
 \hline
  \cite{liu2019heterogeneous,wang2013improved} & \checkmark & & \checkmark & \checkmark   \\
 \hline
 \cite{liu2018network,rosenquist2011social,schaefer2011misery}& \checkmark & \checkmark &  & \checkmark \\ 
 \hline
 \cite{lin2019social} & \checkmark & \checkmark & \checkmark &   \\ 
 \hline
\end{tabular}}
\caption{Summary of existing related work compared to our work proposed in this study. A check mark indicates that the given study satisfies the given criterion.}\label{Table:1}

\end{table} 

The contributions of our study are as follows. Before we develop a predictive model of one's mental health, we perform two exploratory data analyses to ensure that developing such a model makes sense. That is, by analyzing a dynamic social network (containing weekly temporal snapshots constructed from the smartphone data from the NetHealth study, Section \ref{sect:method-data}), we first check whether the group of the depressed (anxious) individuals occupies different social network positions (i.e., has different dynamic network features) than the group of the non-depressed (non-anxious) individuals. Here, the two groups are naturally defined in the data based on one's depression/anxiety trait information. Also, here, we examine \textit{network position} differences in the \textit{trait-based} groups. Second, we check whether different individuals who have different evolving network positions also show depression/anxiety trait differences. Here, to identify the groups of individuals with certain network positions, we use network clustering, which places in the same group those individuals who have similar network positions and in different groups those individuals who have dissimilar network positions. So, here, we examine \textit{trait} differences in the \textit{network-based} groups. If in these two analyses we observe network/trait differences in the trait-based/network-based groups, the network structural information \emph{is} well-associated with the trait information and vice versa, and consequently, it makes sense to develop a predictive model of individuals' mental health traits based on their network structural information. Indeed, this is what we observe in both of the analyses. 

So, third, we develop a predictive model of mental health using three features of individuals extracted from the dynamic network: centralities in the dynamic network, dynamic graphlet degree vectors (dynamic GDV) \cite{hulovatyy2015exploring}, and graphlet orbit transitions (GoT) \cite{aparicio2018graphlet}. To evaluate the predictive power of these features, as a proof-of-concept, we train a logistic regression classifier for each feature to predict individuals as depressed or non-depressed (anxious or non-anxious). To answer our key question\textemdash whether using dynamic network data yields more accurate predictions than using static network data\textemdash we model the dynamic network (see above) as a static network in which two nodes are connected if they have an edge in any temporal snapshot of the dynamic network. We extract two features from the static network: centralities in the static network and static graphlet degree vectors (static GDV) \cite{milenkovic2008uncovering}. For fairness, these two features are static counterparts of two of the above dynamic features: centralities in the dynamic network and dynamic GDV, respectively (the third dynamic feature, GoT, has no static counterpart). Also, for fairness, we use the same logistic regression classifier as above to evaluate the predictive power of the static network features. We find that using any of the three dynamic network features outperforms using both of the static network features (as well as using a fairly comparable non-network feature). This confirms the superiority of using dynamic network data over using static network data in the task of mental health prediction.

%%%%%%%%%%%%%%%%%%%%%%%%%%%%%%%%%%%%%%%%%%%%%%%%%%%%%%%%%%%%%%%%%%%%%%%%%%%%%%%%%%%%%%%%%%%%%%%%%%%%%%%%%%%%%%%%%%%%%%%%%%%%%%%%%%%%%%%%%%%%%%%%%%%%%%%%%%%%%%%%%%%%%%%%%%%%%%%%%%%%%%%%%%%%%%%%%%%%%%%%%%%%%%%%%%%%%%%%%%%%%%%%%%%%%%%%%%%%%%%%%%%%%%%%%%%%%%%%%%%%%%%%

\section{Methods}
\subsection{Dynamic network}
\label{sect:method-data}
\noindent \textbf{Data source.} \hspace{1mm} Our data come from the NetHealth study, which collected smartphone, wearable sensor (Fitbit), and survey-based trait data about $\sim$700 undergraduate student participants at the University of Notre Dame from 2015 to 2019 \cite{purta2016experiences, faust2017exploring}. For reasons stated in Section \ref{intro}, of all NetHealth data, we focus on individuals' SMS logs (i.e., social interactions) collected through smartphones and their mental health trait data on depression and anxiety collected through surveys. Of all participants, 615 were iPhone users and the rest were Android users. We focus on the 615 iPhone users because of issues with the Android data consistency.

\noindent \textbf{Selection of the study time period and pool of individuals.} \hspace{1mm}  We already produced a collection of manuscripts on various research questions related to various dimensions of the NetHealth data \cite{liu2019heterogeneous,liu2018network}. This study is a novel addition to that collection. To be able to draw conclusions across the different dimensions, we match the social interaction and mental health trait data  closely between this current study and our previous studies\cite{liu2019heterogeneous,liu2018network}. This includes the choices of the study time period and pool of individuals, as follows. During the entire 2015-2019 period, as time went on, more and more students dropped out of the study or became data non-compliant. We wish to consider a time period that is as long as possible and that also includes as many individuals as possible; these two conditions conflict with each other. As a result, we focus on the period from August 2015 to August 2016 because only during this period, the vast majority of NetHealth participants were actively involved in the study \cite{liu2019heterogeneous,liu2018network}. This period covers 31  school weeks (we do not consider 21 break weeks, since these do not have meaningful social network structures) \cite{liu2019heterogeneous,liu2018network}. We observe that during our study time period, 576 out of the 615 iPhone users actively sent or received SMSs. We use the 576 individuals' SMS logs (i.e., social interactions) to construct a dynamic social network. For each temporal snapshot of the dynamic network, one snapshot per week, nodes are the individuals and there is an edge between two nodes if there is at least one SMS event between the two corresponding individuals during the given week. The 31 resulting snapshots form the dynamic network.

Of the 576 individuals, 274 are compliant enough and also have mental health trait data  \cite{liu2019heterogeneous,liu2018network}. These 274 individuals form the final pool of individuals to be used in our predictive tasks. That is, we use the  social interaction data of all 576 individuals' to compute each node's features (needed for our predictive tasks) from the entire network. But when we make mental health predictions and evaluate prediction accuracy, we are able to do so only for the 274 individuals for whom we have the mental health data. Of the 274 individuals, 67 individuals (24.5\%) are depressed, and 106 individuals (38.7\%) are anxious (Supplementary Section S1).

%More details on the mental health trait data are as follows. The individuals' depression and anxiety scores are collected from the survey taken in January 2016. We use these scores to divide the 274 individuals into depressed and non-depressed (anxious and non-anxious). We consider individuals whose depression scores, measured by the Center for Epidemiological Studies Depression Scale (CES-D) test, are above 15 as depressed, which is suggested in the literature \cite{kohout1993two,jackson2010race,rosenquist2011social}. Out of the 274 individuals, 67 individuals (24.5\%) are depressed according to this criterion. We consider individuals whose anxiety scores, measured by the State-Trait Anxiety Inventory (STAI) test, are above 40 as anxious, which is suggested in the literature \cite{grant2008maternal,julian2011measures,faisal2007prevalence}. Out of the 274 individuals, 106 individuals (38.7\%) are anxious according to this criterion.

\vspace{-5mm}
\subsection{Network analysis methods}

We perform three tasks. Tasks 1 and 2 rely on centralities of nodes, and task 3 relies on centralities plus other network features. We first discuss the considered network centrality measures, followed by tasks 1, 2, and 3, where the latter covers the other considered features.

\noindent \textbf{Measuring positions (centralities) of nodes in the dynamic network.}

%% centralities
\noindent In each network snapshot, we measure network positions of all nodes with respect to eight popular centrality measures: eccentricity, closeness, betweenness, eigenvector, $k$-coreness, clustering coefficient, degree, and graphlet degree centrality (Supplementary Section S2) \cite{liu2018network,milenkovic2011dominating}. We use the multiple centrality measures because they capture the importance of a node in a network from different perspectives. For each centrality measure, for each node, we compute the given node's centrality value in each snapshot, resulting in 31 centrality values for the 31 snapshots of the dynamic network. Because (\emph{i}) the network snapshots can have different sizes, %(not all nodes are involved in SMS interactions in all snapshots and also different snapshots have different numbers of edges), 
(\emph{ii}) centrality measures can be dependent on the network size, and (\emph{iii}) we aim to study changes in a given node's network position with time (i.e., across the snapshots), we do not consider the ``raw'' centrality values. Instead, we convert these into centrality ranks by giving the individual with the lowest ``raw'' centrality value a rank of 1 (the $1^{st}$ least central), the individual with the next lowest ``raw'' centrality value a rank of 2 (the $2^{nd}$ least central), and so on. We assign the ranks in this way so that the intuition remains the same as with the ``raw'' centrality values: the higher the centrality rank of a node, the more central (i.e., topologically important) the node is. Henceforth, by centrality values, we mean centrality ranks.

\noindent \textbf{Task 1: Do depressed and non-depressed (anxious and non-anxious) individuals occupy different social network positions?}
\label{sect:method-difference}

%\noindent We measure the difference between evolving  centralities of the depressed (anxious) individuals and evolving  centralities of the non-depressed (non-anxious) individuals in two  ways.

\noindent {\fontfamily{cmtt}\selectfont a. Magnitudes of individuals' centralities.} \hspace{1mm} Here, we aim to measure whether centralities of depressed (anxious) individuals are higher or lower (on average over time) than those of non-depressed (anxious) individuals.
%(without necessarily accounting for fluctuation of centralities over time). 
For each centrality measure and each individual, we average the given individual's centralities over the 31 network snapshots. Then, we compare the distributions of the average centralities of the depressed and non-depressed (anxious and non-anxious) individuals using the Wilcoxon rank-sum test, whose $p$-value  quantifies the significance of the difference between the two distributions. Since we do eight tests for the eight centrality measures, we adjust the \textit{p}-values via false discovery rate estimation to correct for the multiple tests \cite{noble2009does}. Throughout this paper, we use the adjusted \textit{p}-value threshold of 0.05. 

\noindent {\fontfamily{cmtt}\selectfont b. Fluctuations of individuals' centralities.} \hspace{1mm} Here, we aim to measure whether centralities of depressed (anxious) individuals vary (increase or decrease) over time more or less than those of non-depressed individuals.
%(without necessarily accounting for the magnitude of centralities). 
For each centrality measure and each individual, we measure the coefficient of variation (CV) of the given individual's centralities in the 31 network snapshots. CV is the ratio of the standard deviation over the average of the 31 centrality values. It is widely used for comparing variability between different samples that have different averages \cite{abdi2010coefficient}. Then, we compare the distributions of CV scores of depressed and non-depressed (anxious and non-anxious) individuals using the Wilcoxon rank-sum test as discussed above.

\noindent \textbf{Task 2: Do individuals who have different evolving network positions show depression/anxiety trait differences?} \label{sect:method-clustering}

\noindent To answer this, we use network clustering to place in the same group (i.e., cluster) those individuals who have similar evolving centrality profiles and in different clusters those individuals who have dissimilar evolving centrality profiles. For reasons described in Supplementary Section S3, we use $k$-medoids clustering under the Euclidean distance and report results for $k=4$, i.e., when obtaining four clusters.
%As a proof of concept (network clustering is not a key focus of this study, and any meaningful choice of network clustering strategy will suffice), for a given centrality measure and an individual $i$, we capture $i$'s 31 centrality values at the 31 network snapshots into $i$'s 31-dimensional vector $V_i$. Then, we compute the dissimilarity between a pair of nodes $i$ and $j$ by calculating the Euclidean distance, a common distance (or equivalently similarity) measure, between $V_i$ and $V_j$. We use a popular clustering method, $k$-medoids clustering, to partition the nodes into groups (i.e., clusters) \cite{kaufman1990partitioning}. The $k$-medoids clustering method requires specifying the number of desired clusters, $k$. To test the effect of $k$, we vary $k$ from 2 to 10 (a relatively large number) in increments of 1. 
%
If the clusters are meaningful, some of them should contain a significant portion of depressed (anxious) individuals while others should contain a significant portion of non-depressed (non-anxious) individuals. We quantify the significance of enrichment of each cluster in depressed and non-depressed (anxious and non-anxious) individuals using the hypergeometric test. Since we  test multiple clusters, we adjust the \textit{p}-values as above. 
%In this study, for illustration purpose, we report results for $k=4$ because using this choice results in the strongest signal in terms of the enrichment significance among all $k$ values that we consider.

\noindent \textbf{Task 3: Is using dynamic network data more accurate than using static network data in predicting mental health?} \label{sect:method-prediction}

\noindent This is our key task. Recall from Section \ref{intro} that we create a static version of our dynamic network.  Then, we compare node features extracted from the dynamic versus static network under the same classifier (see below). We use the following features of a node in a network.

\noindent {\fontfamily{cmtt}\selectfont a. Three dynamic network features.} \hspace{1mm} 

\noindent \textbf{1.} \textit{\underline{Centralities in the dynamic network}}. \hspace{1mm} Recall that we consider eight centrality measures, and that for each measure, we obtain a 31-dimensional vector for each node. To hopefully benefit from the different centrality measures, we integrate their eight 31-dimensional feature vectors, resulting in a final $8 \times 31=248$-dimensional feature vector for each node. \hspace{1mm} \textbf{2.} \textit{\underline{Dynamic GDV}} \cite{hulovatyy2015exploring}. \hspace{1mm} Graphlets, as originally defined in the context of a static network, are small connected non-isomorphic induced subgraphs of such a network \cite{prvzulj2004modeling,milenkovic2008uncovering}. As an extension of static graphlets to the dynamic network setting, dynamic graphlets were introduced by adding temporal information onto edges of a graphlet, which now become events that appear in a certain temporal order. Dynamic GDV of a node characterizes how the extended network neighborhood of the  node evolves by counting for each dynamic graphlet type the number of times the node participates in the given graphlet type. \hspace{1mm} \textbf{3.} \textit{\underline{GoT}} \cite{aparicio2018graphlet}. \hspace{1mm} For a given node, this feature counts how many times in the node's extended network neighborhood each static graphlet type (e.g., a 3-node path) transitions into every other static graphlet type (e.g., a triangle) between every pair of consecutive temporal snapshots. Dynamic GDV and GoT are complementary dynamic network features \cite{aparicio2019temporal}.

\noindent {\fontfamily{cmtt}\selectfont b. Two static network features.} \hspace{1mm} 

\noindent \textbf{1.} \textit{\underline{Centralities in the static network}}. \hspace{1mm} In the static network, for each node, for each of the eight centrality measures, we obtain a single centrality value. To benefit from the different measures, we integrate their eight centrality values, resulting in a final 8-dimensional node feature vector. This is the static counterpart of the centralities in the dynamic network above. \hspace{1mm} \textbf{2.} \textit{\underline{Static GDV}} \cite{milenkovic2008uncovering}. \hspace{1mm} Static GDV characterizes the structure of a node's extended neighborhood in a static network by counting for each static graphlet type the number of times the node participates in the given graphlet type. Static GDV is the static counterpart of dynamic GDV above. Note that there is no static counterpart of GoT.

\noindent {\fontfamily{cmtt}\selectfont c. Raw SMS feature}. \hspace{1mm}  For an individual, we count how many SMSs the  individual sent or received in each of the 31 weeks, which results in a 31-dimensional node feature vector. This is as similar as possible \textit{non-network} counterpart of the centralities in the dynamic network. 
%We use this non-network feature to fairly evaluate the predictive power of its network equivalent.

\noindent {\fontfamily{cmtt}\selectfont d. Feature dimensionality reduction}. \hspace{1mm} When building predictive models, high-dimensional features tend to cause overfitting, meaning that models using such features may fit well on the training data but not predict well on the testing data. Therefore, for each of the features considered above, we generate the corresponding new lower-dimensional feature using principal component analysis (PCA). So, for each of the three dynamic network features, two static network features, and the one non-network feature, we have its pre- and post-PCA versions. Thus, in total, we consider $2 \times (3+2+1)=12$ features.

\noindent {\fontfamily{cmtt}\selectfont e. Classification}. \hspace{1mm}  For each of the 12 features, we train a logistic regression classifier (we use logistic regression as a proof-of-concept), resulting in 12 classification models for predicting an individual as either depressed or non-depressed (anxious or non-anxious). We consider an additional predictive model -- the most accurate HIN-based RS model from our previous work (Section \ref{intro}) \cite{liu2019heterogeneous}, DMF\cite{drumond2014optimizing}. The above dynamic network-based classification models need to be superior to DMF for it to make sense to incorporate in the future the dynamic network data into the HIN-based framework, per the discussion in Section \ref{intro}. Note that although DMF could take an HIN as input \cite{liu2019heterogeneous}, in this study, to fairly compare DMF against the above classification models, we use as DMF's input the same (homogeneous) social network from which we extract the above static network features. In addition, to evaluate the statistical significance of all models' predictive results, we compare them against a random guess model, which works as follows. Recall that in our data, 67 of the individuals are depressed and 207 are non-depressed (106 are anxious and 168 are non-anxious). To make predictions, the random guess model randomly chooses 67 (106) of all individuals and predicts them as depressed (anxious), and it predicts the remaining individuals as non-depressed (non-anxious).

To evaluate the performance of our considered predictive models, we use 5-fold cross-validation (Supplementary Section S4). Given a model's prediction for an individual, taking depression as an example, a true positive (TP) is an individual who is depressed and is also predicted as depressed. A false positive (FP) is an individual who is non-depressed but is predicted as depressed. A false negative (FN) is an individual who is depressed but is predicted as non-depressed. A true negative (TN) is an individual who is non-depressed and is also predicted as non-depressed. Based on these, we compute four popular evaluation measures: precision, recall, F1 score, and accuracy (Supplementary Section S4). 

When comparing performance of any two predictive models in terms of their five paired runs of 5-fold cross-validation, we evaluate the statistical significance of their performance difference by using the Wilcoxon signed-rank test. Since for each model we compare its performance against the rest of the considered models, we adjust the \textit{p}-values as described above.

%We find that for both depression and anxiety, only for the feature \textit{centralities in the static network}, using pre-PCA version is more accurate, and for the rest of the features, using post-PCA is more accurate in terms of all evaluation measures. In this study, for each feature, we consider the better of its pre- and post-PCA versions.   

\section{Results and discussion}

\subsection{Task 1: Depressed and non-depressed (anxious and non-anxious) individuals occupy different social network positions} \label{sect:result-difference}

%We find that depressed individuals have significantly (adjusted \textit{p}-value$<$0.05) lower degree centralities than non-depressed individuals (Figure \ref{fig:1}). Other centrality measures show consistent results (Supplementary Figures S1 and S2). For anxiety, we observe both qualitatively and quantitatively similar results\textemdash anxious individuals have significantly (adjusted \textit{p}-value$<$0.05) lower centralities than non-anxious individuals (Supplementary Figures S3 and S4). In addition, we find that depressed individuals have significantly (adjusted \textit{p}-value$<$0.05) greater fluctuations (i.e., higher CV) of their degree centralities than non-depressed individuals (Figure \ref{fig:1}). Other centrality measures show consistent results (Supplementary Figures S1 and S5). This indicates that social network positions of depressed individuals are more likely to change with time than non-depressed individuals. For anxiety, we find both qualitatively and quantitatively similar results\textemdash the anxious individuals have significantly (adjusted \textit{p}-value$<$0.05) greater fluctuations of their centralities than non-anxious individuals (Supplementary Figures S3 and S6). 

%\vspace{-10mm}

\begin{figure}[h!]
\centering
\includegraphics[width=6.3cm]{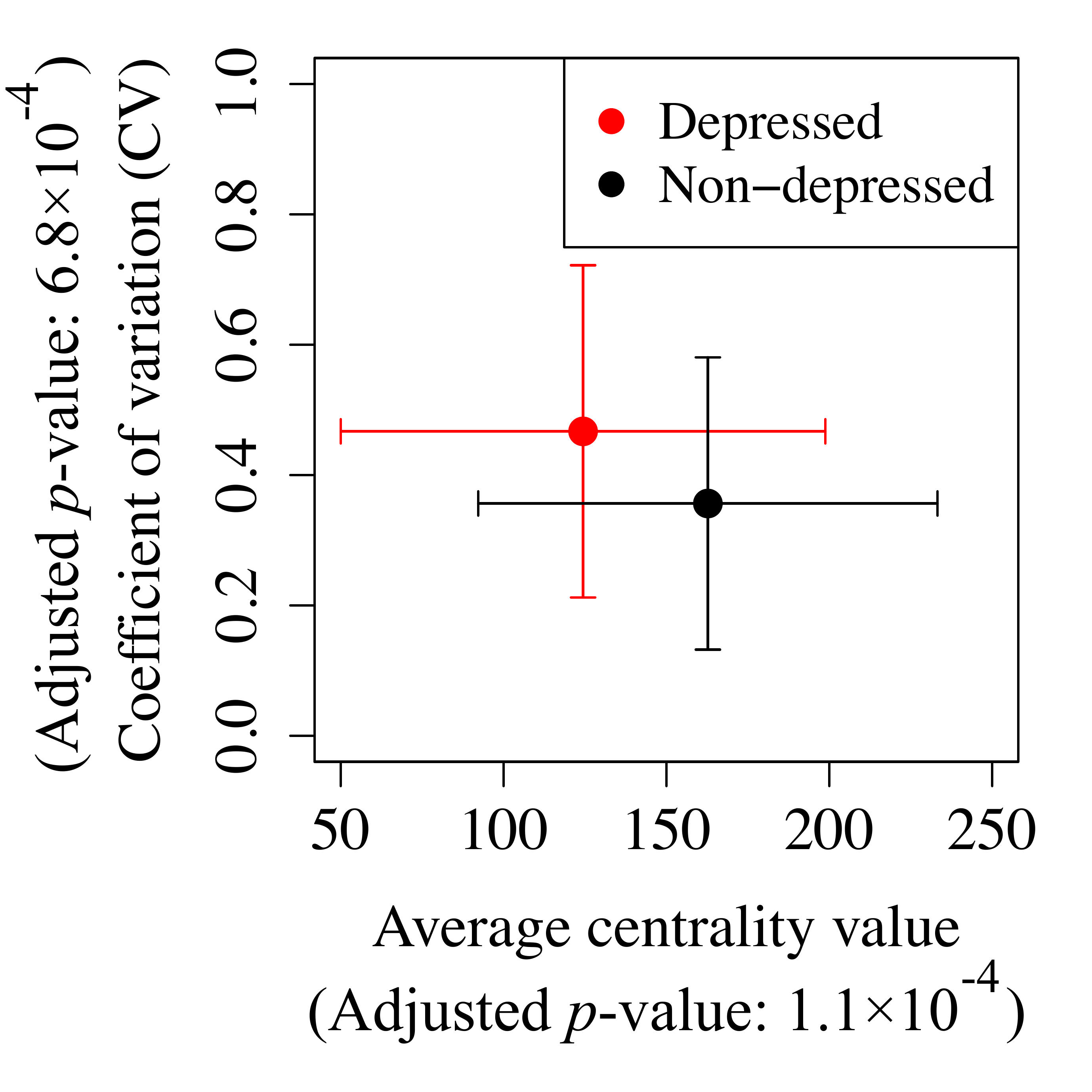}
{\caption{The difference between centralities of the depressed and non-depressed individuals, in terms of centrality magnitudes ($x$-axis) and fluctuations ($y$-axis); see Section \ref{sect:method-difference}. For each of these two quantities, the average over all depressed (red) or non-depressed (black) individuals is shown, along with the corresponding standard deviation (error bars); the adjusted $p$-value of the difference is also shown. This figure is for degree centrality and depression. Results are qualitatively similar for the other centralities and  anxiety (Supplementary Figs. S1-S6). }\label{fig:1}}
\end{figure}

We find that the depressed (anxious) individuals have significantly lower magnitudes of centralities and higher fluctuations of centralities than the non-depressed (non-anxious) individuals (Fig. \ref{fig:1} and Supplementary Figs. S1-S6). The former means that the depressed (anxious) individuals are less central, i.e., are more peripheral, in the social network than the non-depressed (non-anxious) individuals. For example, in terms of degree centrality, the depressed (anxious) individuals have fewer social contacts than the non-depressed (non-anxious) individuals. This result obtained from our social interaction (SMS) data collected via smartphones is consistent with existing knowledge that depression and anxiety are associated with having fewer friends based on social interaction data collected from surveys \cite{alpass2003loneliness,george1989social,schaefer2011misery}, which validates that smartphone data may be a good proxy for real-world friendships. The latter means that centralities of the depressed (anxious) individuals vary more with time than centralities of the non-depressed (non-anxious) individuals. For example, in terms of degree centrality, the depressed (anxious) individuals have many friends at some time points but few friends at other time points, while the non-depressed (non-anxious) individuals have more stable friendships. This result indicates the promise of using dynamic network data over using static network data in the task of predicting mental health, since the latter might fail to capture such temporal variation. 

Since the depressed and non-depressed (anxious and non-anxious) individuals have different network positions, it makes sense to use supervised learning methods to predict the individuals' mental health based on their network positions, which we do in task 3 below.

%\vspace{-0.1cm}

\vspace{-0.9cm}

\subsection{Task 2: Individuals who have different evolving network positions show depression/anxiety trait differences}\label{sect:result-clustering}

\begin{figure}[h!]
\centering
\includegraphics[width=6cm]{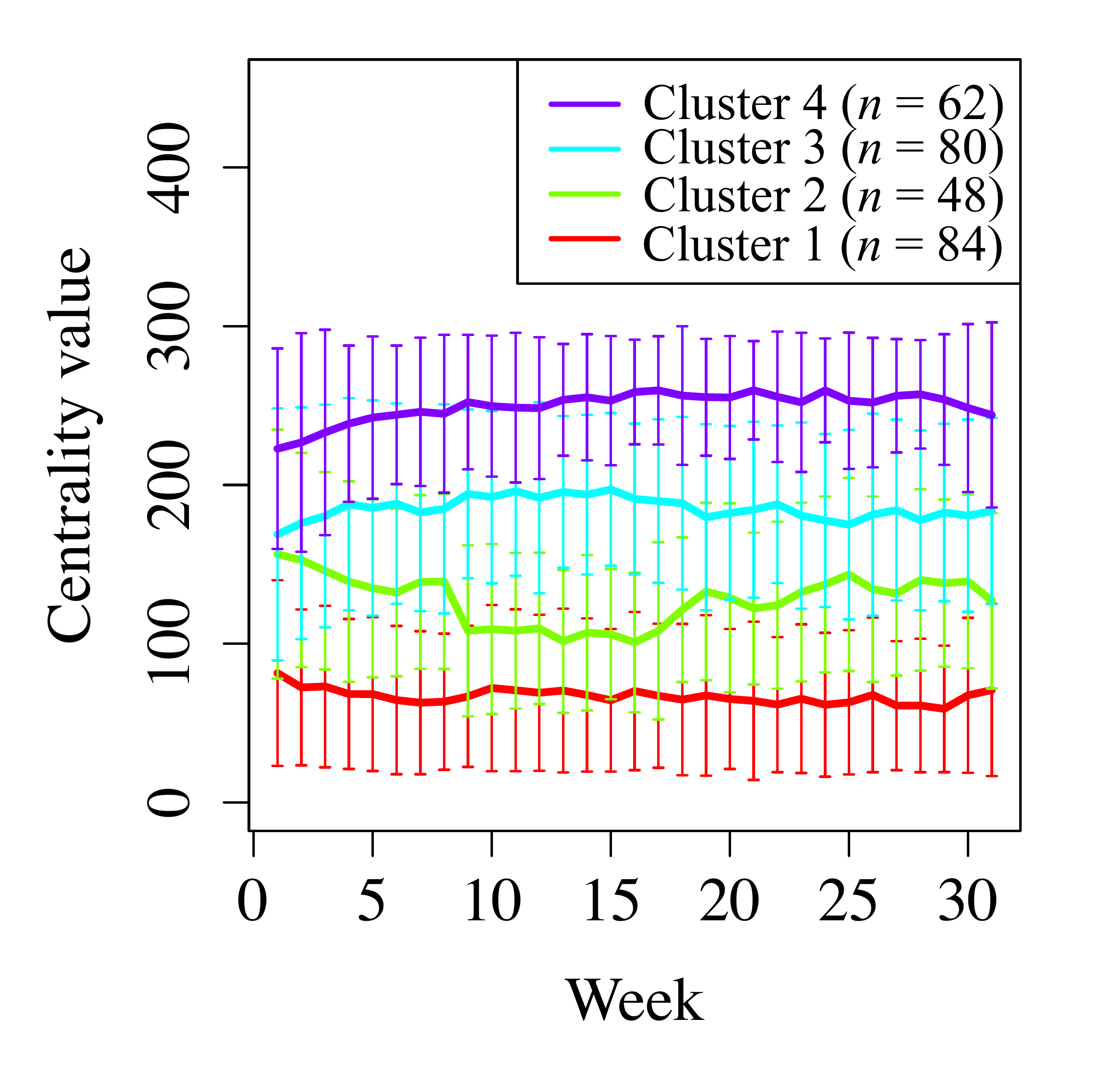}
{\caption{Centrality values of individuals in the four network-based clusters. For a given cluster, at each time point (week), the average centrality value over all individuals in the cluster is shown, along with the corresponding standard deviation (error bars); also, the number of individuals in the cluster (\textit{n}) is shown. Clusters are ordered from the one with the lowest centralities (red) to the one with the highest centralities (purple). This figure is for degree centrality-based clusters. Results are qualitatively similar for the other centralities.
}\label{fig:2}}

\end{figure}

\begin{figure}[h!]
\hspace{-0.25cm}
\begin{subfigure}{0.502\linewidth}
\includegraphics[width=\linewidth]{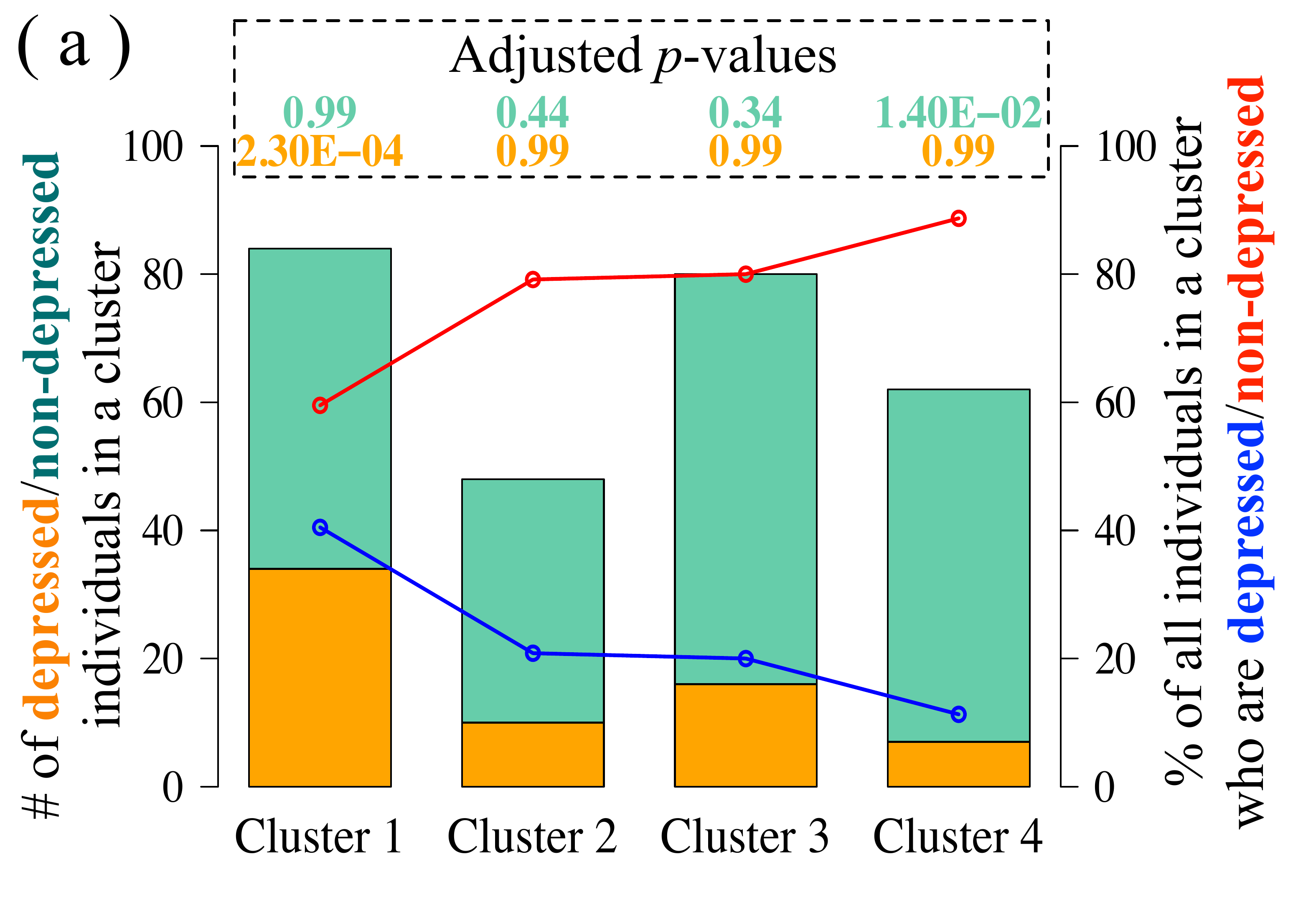}
\subcaption{}
\label{fig:3}
\end{subfigure}
\hfill 
\hspace{-0.21cm}
\begin{subfigure}{0.502\linewidth}
\includegraphics[width=\linewidth]{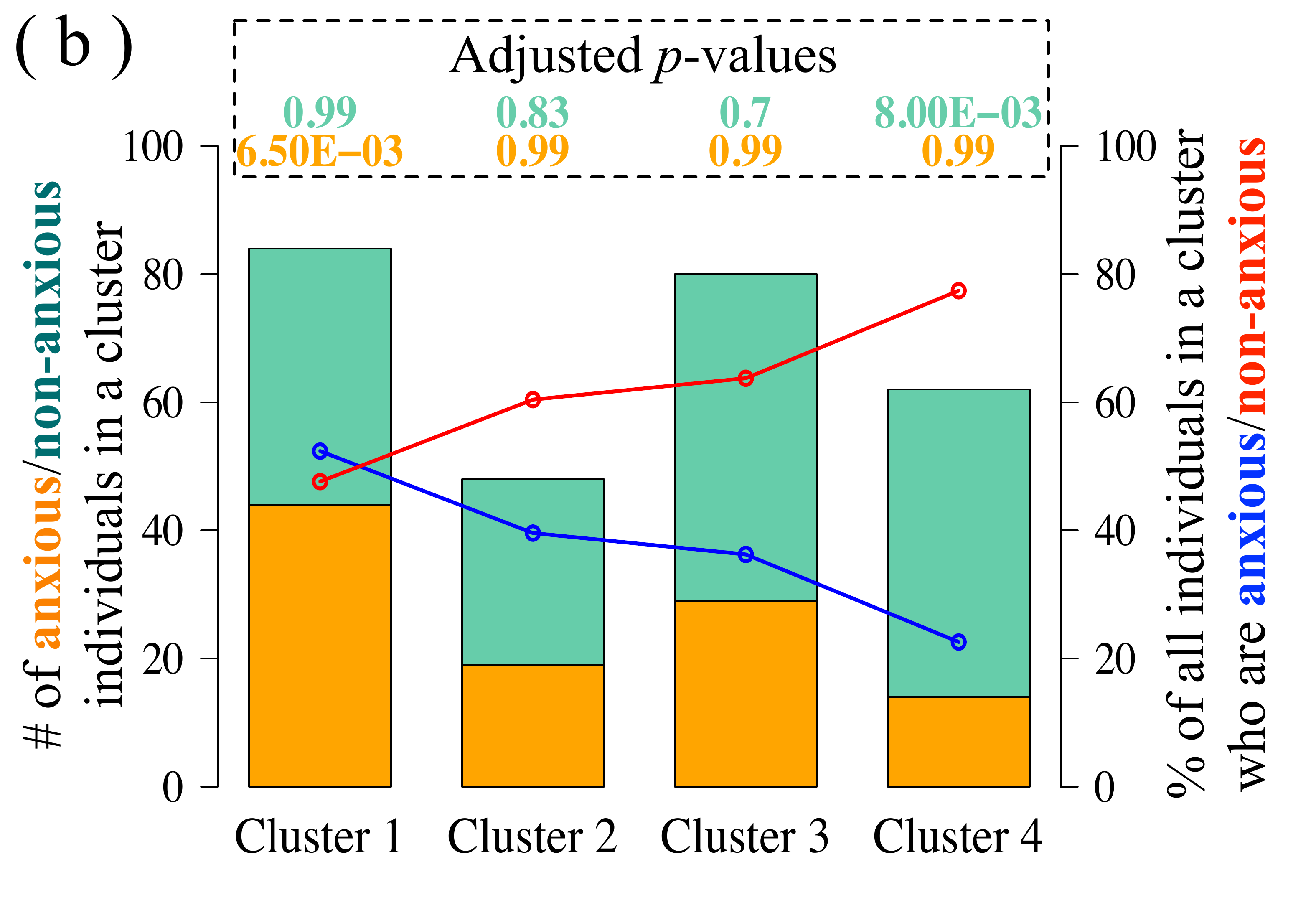}
\subcaption{}
\label{fig:4}
\end{subfigure}
\vspace{-0.9cm}
\caption{The enrichment of the four network-based clusters in \textbf{(a)} depressed and non-depressed individuals, and \textbf{(b)} anxious and non-anxious individuals. In a given panel, the left $y$-axis shows the number of depressed and non-depressed (anxious and non-anxious) individuals in a given cluster. For example, in panel (a), in cluster 1 containing 84 individuals, there are 34 depressed and 50 non-depressed individuals. The right $y$-axis shows the percentage of depressed and non-depressed (anxious and non-anxious) individuals in a given cluster. For example, in panel (a), of the individuals in cluster 1, 40\% are depressed and 60\% are non-depressed (the two percentages add up to 100\%). On top of a given bar (corresponding to a given cluster), the adjusted \textit{p}-values are shown indicating the statistical significance of the enrichment of the cluster in depressed and non-depressed (anxious and non-anxious) individuals. For example, in panel (a), cluster 1 is significantly enriched (adjusted $p$-value$<$0.05) in depressed individuals, and cluster 4 is significantly enriched in non-depressed individuals; no other cluster is significantly enriched in either depressed or non-depressed individuals (this is directly related to the fact that the blue values corresponding to the enrichment in depressed individuals decrease going from cluster 1 to cluster 4, and because the red values corresponding to  the enrichment in  non-depressed individuals decrease going from cluster 4 to cluster 1). These results are for degree centrality-based clusters. Results are qualitatively similar for the other centralities.}\label{fig:3entire}
\end{figure}

We find that the individuals in the different network-based clusters (Fig. \ref{fig:2}) have different depression (anxiety) traits. Namely, the lower the centrality values (cluster 1 being the least-central), the more likely that individuals are depressed (anxious), and the higher the centrality values (cluster 4 being the most-central), the more likely that individuals are non-depressed (non-anxious)  (Fig. \ref{fig:3entire}). While we illustrate these results for degree centrality, results  are qualitatively similar for the other centrality measures (not shown due to space constraints). 

Thus, network-based features of individuals (their evolving centrality profiles) can  distinguish well between depressed (anxious) and non-depressed (non-anxious) individuals. This is an additional confirmation that it makes sense to develop a predictive model of individuals' mental health based on their network structural information, which is what we do next.

%\vspace{-1cm}
\vspace{-0.8cm}

\subsection{Task 3: The dynamic network has more power to predict individuals' mental health than the static network} \label{sect:result-prediction}

Recall from Section \ref{sect:method-prediction} that we develop a predictive classification model for each of the three dynamic and two static network features, plus the dynamic raw SMS data (non-network) feature. For each feature, we focus on the best of its pre- and post-PCA versions; we have found that using the post-PCA version is more accurate than using the pre-PCA version for each considered feature except one, namely centralities in the static network. Also, we evaluate the static RS model, DMF, and the random guess model. We evaluate the  models with respect to four evaluation measures: precision, recall, F1 score, and accuracy.

\begin{figure}[h!]
\centering
\begin{subfigure}[b]{0.49\columnwidth}
\includegraphics[width=\linewidth]{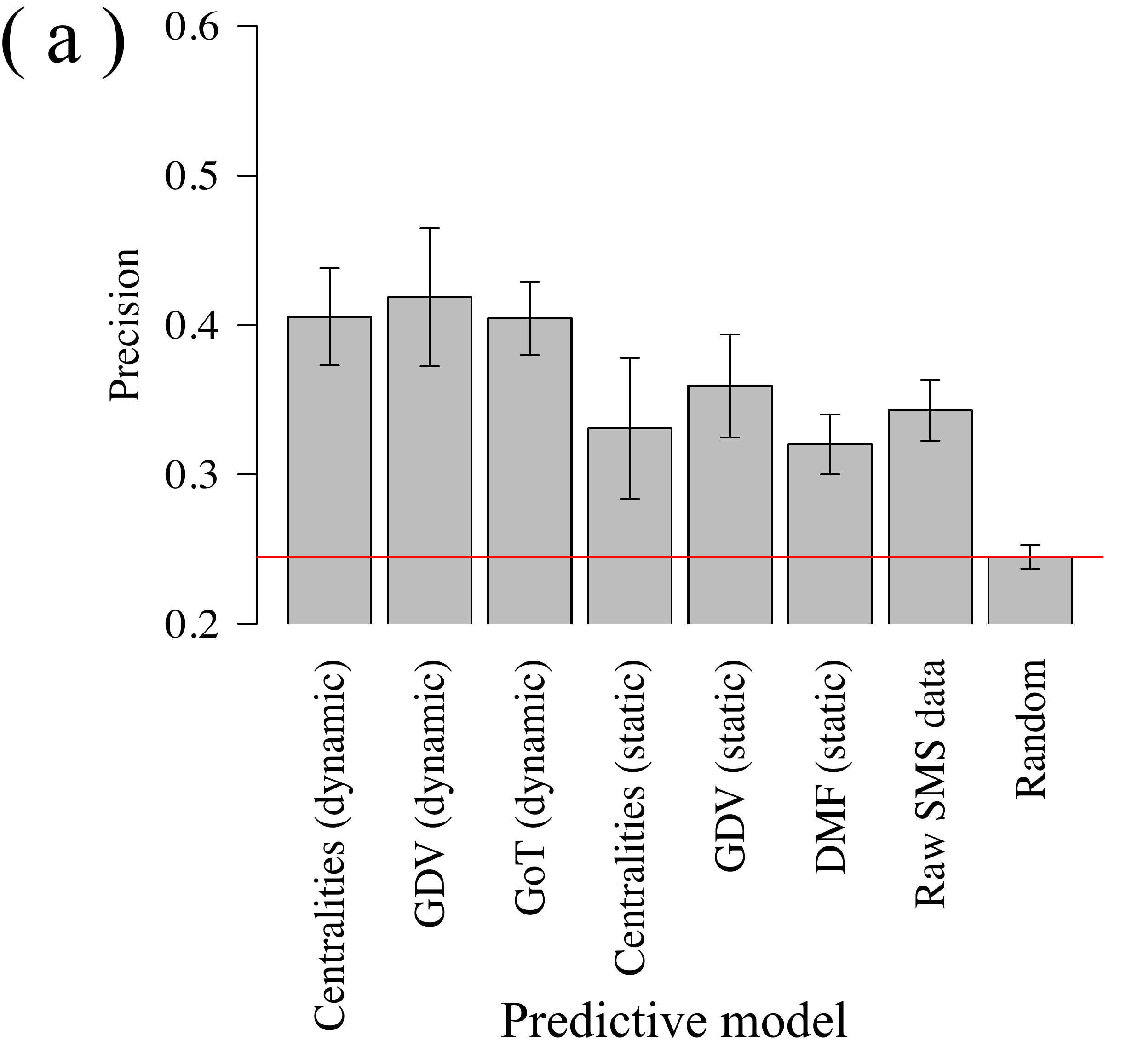}
\subcaption{}
\label{fig:5}
\end{subfigure}
\hfill
\begin{subfigure}[b]{0.49\columnwidth}
\includegraphics[width=\linewidth]{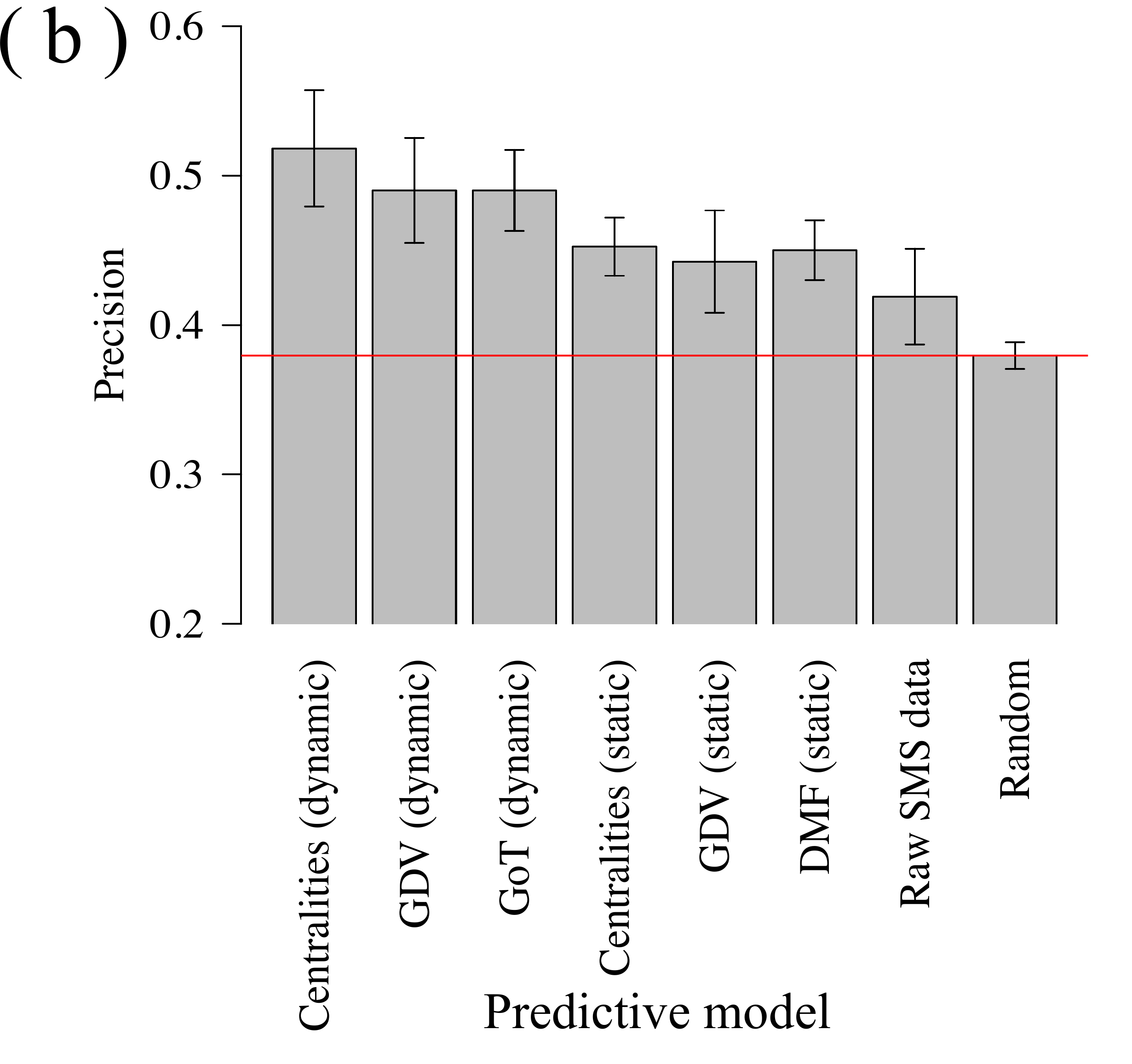}
\subcaption{}
\label{fig:6}
\end{subfigure}
\vspace{-0.85cm}
\caption{The performance of the considered predictive models for \textbf{(a)} depression and \textbf{(b)} anxiety. For each model, the  performance averaged over the five runs of 5-fold cross-validation is shown, along with the corresponding standard deviation. In parentheses, we show whether the feature in a given model is extracted from the dynamic or static network. The red line corresponds to the performance of the random guess model. These results are for precision. Results are qualitatively similar for the other evaluation measures (Supplementary Figs. S7-S8).}\label{fig:4full}
\end{figure}

Our findings are as follows, and they hold for both depression and anxiety, as well as for   all four evaluation measures. All three dynamic network feature-based classification models are significantly more accurate (adjusted \textit{p}-value$<$0.05)  than (\emph{i}) both of the static network feature-based classification models, (\emph{ii}) the DMF model, (\emph{iii}) the non-network classification model, and (\emph{iv}) the random guess model (Fig. \ref{fig:4full} and Supplementary Figs. S7-S8). This confirms the hypothesis of our paper that the dynamic network features have more predictive power than any one of the other features. The three dynamic network feature-based classification models perform similarly, with none of them having perfect performance. This indicates their likely complementarity and a potential promise of developing in the future an ensemble learning approach that would integrate the different dynamic network features.

%%%%%%%%%%%%%%%%%%%%%%%%%%%%%%%%%%%%%%%%%%%%%%%%%%%%%%%%%%%%%%%%%%%%%%%%%%%%%%%%%%%%%%%%%%%%%%%%%%%%%%%%%%%%%%%%%%%%%%%%%%%%%%%%%%%%%%%%%%%%%%%%%%%%%%%%%%%%%%%%%%%%%%%%%%%%%%%%%%%%%%%%%%%%%%%%%%%%%%%%%%%%%%%%%%%%%%%%%%%%%%%%%%%%%%%%%%%%%%%%%%%%%%%%%%%%%%%%%%%%%%%%

\vspace{-0.7cm}

\section{Conclusions}
In this paper, we develop a predictive model of mental health that uses rich dynamic social network data. We demonstrate that using the dynamic network data has advantage over using its static network equivalent as well as its non-network equivalent in this predictive task. As a consequence, our previous study that is the most comprehensive work to date on predicting individuals' mental health in terms of  heterogeneity of the considered data  (including e.g., Fitbit data)\cite{liu2019heterogeneous}, which yielded high prediction accuracy despite using static social network data (which had to be used due to current methodological challenges in the field of heterogeneous network analysis), could be further improved by incorporating the data dynamics. This  non-trivial direction that requires novel algorithmic thinking is the subject of our future work.

%%%%%%%%%%%%%%%%%%%%%%%%%%%%%%%%%%%%%%%%%%%%%%%%%%%%%%%%%%%%%%%%%%%%%%%%%%%%%%%%%%%%%%%%%%%%%%%%%%%%%%%%%%%%%%%%%%%%%%%%%%%%%%%%%%%%%%%%%%%%%%%%%%%%%%%%%%%%%%%%%%%%%%%%%%%%%%%%%%%%%%%%%%%%%%%%%%%%%%%%%%%%%%%%%%%%%%%%%%%%%%%%%%%%%%%%%%%%%%%%%%%%%%%%%%%%%%%%%%%%%%%%

\section{Acknowledgements}
This work was funded by the National Institutes of Health (NIH) 1R01HL117757 and National Science Foundation (NSF) CAREER CCF-1452795 grants.

\bibliographystyle{abbrv}
\bibliography{main}

\end{document}